# Structure and dielectric dispersion in cubic-like $0.5K_{0.5}Na_{0.5}NbO_3$-$0.5Na_{0.5}Bi_{0.5}TiO_3$ ceramic

Laijun Liu[1,2*], Michael Knapp[1], Ljubomira Ana Schmitt[3], Helmut Ehrenberg[1], Liang Fang[2], Hartmut Fuess[3], Markus Hoelzel[4], Manuel Hinterstein[1,5]

[1]Karlsruhe Institute of Technology (KIT), Institute for Applied Materials (IAM), Hermann-von-Helmholtz-Platz 1, D-76344 Eggenstein-Leopoldshafen, Germany

[2]College of Materials Science and Engineering, Guilin University of Technology, Guilin 541004, China

[3]Institute of Materials and Geoscience, Technische Universität Darmstadt, Alarich-Weiss-Straße 2, Darmstadt 64287, Germany

[4]Forschungsneutronenquelle Heinz Maier-Leibnitz (FRM II), Technische Universität München, Lichtenbergstrasse 1, D-85747 Garching, Germany

[5]UNSW Australia, School of Materials Science and Engineering, 2052 Sydney, Australia

## Abstract

The nature of the cubic-like state in the lead-free piezoelectric ceramics $0.5K_{0.5}Na_{0.5}NbO_3$-$0.5Na_{0.5}Bi_{0.5}TiO_3$ (KNN-50BNT) has been examined in detail by synchrotron x-ray diffraction (SD), selected area electron diffraction (SAED), neutron diffraction (ND), and temperature dependent dielectric characterization. The SD pattern of KNN-50BNT presents a pure perovskite structure with pseudocubic symmetry. However, superlattice reflections were observed by SAED and completely indexed by tetragonal symmetry with *P*4*bm* space group in ND pattern. The relaxor behavior of KNN-50BNT is compared with Pb-based and Ba-based relaxors and discussed in the framework of the Vogel-Fulcher law and the *new glass model*. The KNN-50BNT ceramic exhibits the strongest dielectric dispersion among them.

---

* Author to whom correspondence should be addressed. Electronic mail: ljliu2@163.com.

**I. Introduction**

Lead-free relaxor materials are of great interest for both environmental protection and fundamental studies. To extend the materials applicability to modern electronics, more detailed information on local structure and physical properties are necessary[1,2]. The local structure of $K_{0.5}Na_{0.5}NbO_3$- (KNN)[3,4] and $Bi_{0.5}Na_{0.5}TiO_3$ (BNT)-based[5-8] materials is interesting.

There is unanimity with regard to the importance of aliovalent substitution for achieving a broad peak of the dielectric permittivity in KNN, e. g. most of the previous work is focused on (1-x)KNN-xBNT with x =0-0.1[9-11]. However, the local structure-property correlations in KNN/BNT-derived relaxors is far from settled primarily due to the persistence of ambiguity with regard to the cubic-like crystal structure of the solid solutions, especially, at a high degree of substitution.

In this letter, the structure of the midpoint of the binary system BNT-KNN was investigated by synchrotron x-ray diffraction (SD), selected area electron diffraction (SAED) and neutron diffraction (ND). A coexistence of cubic and tetragonal structure exhibits anomalously large dielectric dispersion. The dielectric dispersion was analyzed by the Vogel-Fulcher relationship and the new glass model, and compared with archetypal relaxor ferroelectrics.

**II. Experimental Procedure**

$0.5K_{0.5}Na_{0.5}NbO_3$-$0.5Na_{0.5}Bi_{0.5}TiO_3$ (KNN-50BNT) ceramics for x-ray diffraction and electrical measurements were prepared by a high-energy ball-milling method[12] while the ceramic for neutron diffraction was prepared by the conventional solid state route.

The high-energy x-ray diffraction experiments were carried out at the Beamline P02.1 (PETRA III) at HASYLAB (DESY, Hamburg, Germany)[13]. This beamline operates at a fixed energy of ~60 keV. ND was done in the SPODI powder diffractometer at the FRM-II (Garching, Germany) at an incident wavelength of 1.548 Å[14]. Full-profile Rietveld refinement was done using the software package FULLPROF[15]. The TEM experiments were performed using a Philips CM20 Super twin microscope. SAED patterns were recorded. A silver electrode paste was then applied and fired at 700°C for 30 min for dielectric measurements. The temperature dependence of dielectric permittivity and loss factor were measured using a high precision LCR meter (HP 4284A) and an environmental oven with a temperature range from 90 to 500K.

**III. Results and Discussion**

SD pattern of the KNN-50BNT is shown in Fig. 1a, which presents a pure perovskite structure with pseudocubic symmetry. No superlattice reflections or reflection splitting can be observed in the SD. Furthermore, no obvious ferroelectric distortion from cubic symmetry can be observed from the diffraction profiles of $110_c$ and $200_c$ pseudocubic Bragg reflections (inset of Fig. 1a). A $Pm\bar{3}m$ model was employed to refine the pattern. The reliability factors of the Rietveld refinements are: $R_p$=10.1%, $R_{wp}$=10.0% and $R_e$=3.50%. The $R$ values are still high, indicates the local structure could be different from the macrostructure detected by the SD.

SAED patterns were carried out to detect the local structure of KNN-50BNT by examining representative grains on the $<011>_c$ and $<100>_c$ zone axes, as shown in Fig. 1b and 1c, respectively. No reflection splitting can be revealed. Additional reflections are visible in the SAED due to diffraction from neighboring grains. The ½ *ooo* superstructure reflections (Fig. 1b) are invisible in the <011>$_c$ zone, while the ½ *ooe* superstructure reflections (Fig. 1c) are visible in the <001>$_c$ zone, where *e* and *o* denote even and odd Miller indices, respectively. It indicates that the local structure is not consistent with the SD result. According to the Glazer notation[16] the superstructure reflections correspond to the $a^0a^0c^+$ tilt system, which can be modeled within the space group *P4/mbm* (centrosymmetric) or *P4bm*[17] (non-centrosymmetric). It is worth to note that the local structure of KNN-50BNT is different from other morphotropic BNT-based ceramics, such as 0.96BNT-0.06BaTiO$_3$[18], 0.92BNT-0.06BaTiO$_3$- 0.02KNN[19] and 0.90BNT-0.07BaTiO$_3$-0.03KNN[20], which include a rhombohedral *R3c* phase or a cubic $Pm\bar{3}m$ phase beside the tetragonal *P4bm* phase. The weak superstructure reflections in the SAED patterns of KNN-50BNT indicate a low proportion of the tetragonal phase with/or small octahedral tilting angle. Therefore, we used new models including *P4/mbm* or *P4bm* to refine the SD pattern. The lattice parameters were obtained by $Pm\bar{3}m$+*P4/mbm* model *a* = 3.9361(7) Å for $Pm\bar{3}m$ phase and *a*=*b* =5.5668(9) Å, c=3.9362(4)Å for *P4/mbm* phase. The atom positions of *P4/mbm* space group were refined to be Na/K/Bi 2c (0, ½, ½), Ti/Nb 2a (0, 0, 0), O1 2b (0, 0, ½), and O2 4g [0.2913(6), 0.7913(6), 0]. A satisfactory fit with the residuals $R_p$=11.7%, $R_{wp}$=11.8% and $R_e$=3.50% was achieved. However, compared with the above model, the best fitting with $R_p$=8.77%, $R_{wp}$=8.96% and $R_e$=3.42% was obtained by $Pm\bar{3}m$+*P4bm* model. The associated refined structural parameters are shown in Table 1. The latter model, therefore, is more reasonable to describe the structure. The tetragonal phase with $c/a$=1.000(4) has nearly the same pseudocubic character. It is considered that the KNN-50BNT includes a pseudocubic matrix implanted with non-centrosymmetric tetragonal phase. This is very similar to the idea of polar nanoregions.

The high-resolution ND pattern of KNN-50BNT is shown in Fig. 2. In contrast to the SD pattern, it shows additional superlattice reflections suggesting oxygen octahedral tilting[19]. The in-phase tilting of oxygen octahedra results in the ½ *ooe* superlattice reflections at 2$\theta$ near 70°, which is consistent with the SAED results. The superlattice reflections contribute to the $a^0a^0c^+$ tilt system which can be described with the space group *P4bm*. Since XRD is less sensitive to oxygen positions, the superlattice reflections from oxygen octahedral tilting are not clear in the SD. Considering the results of the SD and the SAED, a Rietveld fitting of the entire pattern using the $Pm\bar{3}m$+*P4bm* model was carried out. Fitting results are shown in Table 2. Good match between observed and calculated profiles was gained for both the main reflections for perovskite and the superlattice peaks for octahedral tilting. Comparing with the fitting results of SD, the differences of atom positions and phase fraction are present, which should be related to the preparation method.

The temperature dependence of the real part ($\varepsilon'$) and imaginary part ($\varepsilon''$) of the dielectric permittivity is shown in Fig. 3a and 3b, respectively. A frequency-dependent broad maxima with frequency dispersion is discerned in both $\varepsilon'$ and $\varepsilon''$ for the KNN-50BNT ceramic. The profiles of the frequency dispersion are very similar to PLZT 8/65/35 relaxors that are of cubic symmetry at room temperature[21], but different from 0.94BNT-0.06BaTiO$_3$ ceramics in which strong frequency dispersion appears in the temperature range before the maximum of $\varepsilon$ ($T'_m$)[22]. The relaxation behavior is not caused by space charge polarization or ion hopping because the complex permittivity does not follow the Cole-Cole equation[23], as shown in Fig. 4. Both $\varepsilon'$-$T$ and $\varepsilon''$-$T$ exhibit a broad asymmetric peak shape for KNN-50BNT. The Curie-Weiss law was employed to fit the $T$~$1000/\varepsilon'$ at 10 kHz to get the Burns temperature $T_B$ ~ 520K, which is shown in the inset of Fig. 3. Compared with the curve of $\varepsilon'$ and $\varepsilon''$, the dispersion is more noticeable before the signal crowns in $\varepsilon'$, and the temperature for the maximum $T''_m$ in $\varepsilon''$ is much lower than $T'_m$ in $\varepsilon'$. It indicates that the appearance of $T'_m$ associates with a thermal evolution of relaxation time or correlation length distribution of polar nanoregions (PNRs) due to not any measurable structural transition (see Raman shift of $v_2$ model in the inset of Fig. 3). Therefore, the PNRs could associate with the *P4bm* phase rather than *P4/mbm* (a centrosymmetric space group that allows superstructure reflections but does not allow cation displacements). This is also in agreement with the structure refinement of the SD and ND.

The frequency $\omega$ dependence of $T'_m$ (or $T''_m$) gives information about the dielectric permittivity and the relaxation time of dielectrics. The relaxation time for the relaxation associated with space charge or ion hopping follows the Arrhenius law[23]. In relaxors, an empirical Vogel-Fulcher relationship is often employed to describe the behavior of $\omega$ ($T'_m$ or $T''_m$) given as[24]

$$\omega = \omega_{01} \exp[\frac{-E_{a1}}{k_B(T-T_f)}] \qquad (1)$$

where $T_f$ is the freezing temperature of the dipoles. The relation indicates that the dielectric spectrum becomes infinitely broad at $T_f$ implying the relaxation time constant approaches infinity. In order to characterize the dielectric relaxation strength of relaxors, Cheng et al.[25,26] introduced the "*New glass model*". A superexponential function to describe the relaxation in $T_m$ is given as

$$\omega = \omega_{02} \exp[\frac{-E_{a2}}{k_B T}]^p \qquad (2)$$

where $p$ ($p>1$) is a constant connected to the degree of relaxation of relaxors. The magnitude of $p$ is considered to increase from unity for materials with a Debye medium to infinity for a long-range ferroelectric order. Therefore, the value of $p$ can be seen as a reflection to quantitatively characterize relaxors.

The application of Eq. (1) to the experimental data is shown in Fig. 3c and the fitting results of $\omega_{01}$, $E_a$, and $T_f$ are summarized in Table 3. It can be seen that $E_a$ is consistently about 0.20 eV from fitting by both $\omega$-$T'_m$ and $\omega$-$T''_m$ relations. However, the $T_f$ obtained from $\omega$-$T'_m$ is much higher than that of $\omega$-$T''_m$. The difference, ~100K is close to the shift between $T'_m$ and $T''_m$ (Fig. 3). Fitting the experimental results by Eq. (2) is shown in Fig. 3d, the magnitude of $p$ and $E_a$ can be determined in the vicinity of $T'_m$ (or $T''_m$). The magnitudes of the fitted parameters are listed in Table 3. Similar to the result of Eq. (1), the obtained $p$ value and activation energy $E_a$ is similar, while $\omega_0$ is different. In ferroelectric relaxors, $\omega_{02}$ denotes the size of polar clusters and the interaction among them. The stronger the interaction and/or the larger the correlation length of clusters, the smaller is the magnitude of $\omega_{02}$[29]. Compared with 0.9PMN-0.1PT (the $T_f$ value between $\omega$-$T'_m$ and $\omega$-$T''_m$ is close though $\omega_{01}$s >$10^{14}$ Hz are not physically reasonable), the differences of $T_f$ and $\omega_{02}$ from the $\varepsilon'$ and $\varepsilon''$ for KNN-50BNT could contribute to different mechanisms. Since *P4bm* is a weakly polar (or non-polar) tetragonal phase[30] and no dielectric anomaly can be observed for the phase transition from *P4bm* to $Pm\bar{3}m$ in BNT-based ceramics[6], the relaxation behavior with a frequency dispersion in $\varepsilon'$-$T$ should contribute to the PNR behavior in the cubic matrix rather than the phase transition from tetragonal to cubic. Anyway, compared with the $p$ value with $BaTiO_3$-based and Pb-based relaxors, KNN-50BNT shows the lowest $p$ value (Table 3), which indicates the strongest degree of relaxation. This behavior could associate with the near non-polar *P4bm* phase.

In most KNN solid solution, $T_C$ (or $T_m$) shifts to low temperature while the peak of the dielectric permittivity is depressed and broaden with the increase of the second component. At a high degree of substitution in (1-x)KNN-x$SrTiO_3$ (x ≥ 0.15)[31] and (1-x)KNN-x$Ba_{0.5}Sr_{0.5}TiO_3$ (x ≥ 0.10)[32], the peak shows a frequency dispersion like a relaxor, although the macroscopic symmetry remains pseudocubic below $T_m$. Therefore, in KNN-50BNT, aliovalent substitution on both A- and B- sites gives rise to the local elastic fields and the local charge imbalance then induce quenched random electric field. It prevents long-range ordering and form PNRs[33]. On the other hand, the aliovalent substitutions both in the two sites hinders the onset of the normal ferroelectric state and enhances dielectric relaxation below $T_m$.

## IV. Conclusions

The local structure and relaxor behavior of KNN-50BNT has been reported in this letter. The phase coexistence of a pseudocubic and near non-polar tetragonal phase is revealed. The anomalous dielectric dispersion occurs due to quenched random electric fields induced by aliovalent substitution on both A- and B- sites. The relaxor behavior was analyzed by the Vogel-Fulcher relationship and the *new glass model*. In comparison with the equivalent data from the classical perovskite relaxors, $BaTi_{0.7}Sn_{0.3}O_3$ and 0.9PMN-0.1PT, the investigated system KNN-50BNT shows the strongest degree of relaxation. This behavior should associate with the near non-polar *P4bm* phase and local structure.

**Acknowledgments**

We thank the support from the fellowship of the Helmholtz Institute and the Karlsruhe Institute of Technology. This work has benefited from beamtime allocation at P02.1 at PETRA III in Hamburg (Germany) and SPODI at FRM II (Garching) Germany. Financial support from the 'Bundesministerium für Bildung und Forschung (BMBF)' under grant number 05K13VK1, the Australian Research Council (ARC) under grant number DE150100750 and SFB 595, the Natural Science Foundation of China (Nos. 11264010, 51002036, 11564010) and the Natural Science Foundation of Guangxi (Grant No. GA139008) is also acknowledged.

**Table and Figure captions**

Fig. 1  Rietveld refined synchrotron x-ray data of KNN-50BNT with $Pm\bar{3}m$ model. Pseudocubic $110_c$ and $200_c$ Bragg reflections are shown in the inset. The circles indicate the observed pattern, the solid lines depict the calculated pattern. Selected-area electron diffraction (SAED) patterns of the $[110]_c$ and $[100]_c$ zone axes are shown in (b) and (c), respectively. The arrow indicates the ½*ooe* superstructure reflection. The weak intensity of the ½*ooe* superstructure reflections signifies a low fraction of the tetragonal *P4bm* phase.

Fig. 2  Rietveld refined neutron diffraction data $Pm\bar{3}m$+*P4bm* phase coexistence model. Tetragonal superstructure reflections ½$310_c$, ½$530_c$ and ½$532_c$ are visible in the insets.

Fig. 3  Temperature dependence of the (a) real and (b) imaginary parts of the relative dielectric permittivity of KNN-50BNT ceramic at 1 kHz, 5kHz, 10 kHz, 50kHz, 100kHz, 500kHz and 1 MHz. Inset: Temperature dependence of $v_2$ Raman model shifts (Blue symbol) and 1000/$\varepsilon'$ (black symbol), the solid curve is fitted by Curie-Weiss law. (c) the plots of ln$\omega$ vs $T_m$ and $T_m'$, the solid curves are fitted according to the Vogel-Fulcher relation (Eq. 1). (d) the plots of ln$\omega$ vs 1000/$T_m$ and 1000/$T_m'$, the solid curves are fitted according to the new glass model (Eq. 2).

Fig. 4  Cole-Cole plot of $\varepsilon'$ vs $\varepsilon''$ for the dielectric relaxation near $T_m$ of KNN-50BNT at different temperatures

Table 1  Structural parameter, fractional atomic coordinates and equivalent isotropic displacement parameter from the synchrotron XRD refinements of the sample of KNN-50BNT at 300K. The isotropic displacement parameter B is in Å$^2$

Table 2  Structural parameter, fractional atomic coordinates and equivalent isotropic displacement parameter from the neutron refinements of the sample of KNN-50BNT at 300K. The isotropic displacement parameter $B$ is in Å$^2$

Table 3  Fitting results according to the Vogel-Fulcher relation and the *new glass model*.

Table 1 Structural parameter, fractional atomic coordinates and equivalent isotropic displacement parameter from the synchrotron XRD refinements of the sample of KNN-50BNT at 300K. The isotropic displacement parameter B is in Å$^2$

| x=0.50 at 300K | Tetragonal phase (P4bm) | | | | Cubic phase (Pm-3m) | | | |
|---|---|---|---|---|---|---|---|---|
| a, b, c (Å) | 5.5661(2) | 5.5661(2) | 3.9372(5) | | 3.9358(1) | 3.9358(1) | 3.9358(1) | |
| α, β, γ (°) | 90 | 90 | 90 | | 90 | 90 | 90 | |
| Na/K/Bi x, y, z B | 0 | ½ | 0.5420(4) | 5.0704(3) | 0 | 0 | 0 | 5.7448(9) |
| Nb/Ti x, y, z B | 0 | 0 | 0 | 0.6242(5) | ½ | ½ | ½ | 0.4746(0) |
| O1 x, y, z B | 0 | 0 | 0.5186(1) | 1.8118(2) | ½ | ½ | 0 | 0.7497(3) |
| O2 x, y, z B | 0.2364(8) | 0.2635(2) | 0.0074(9) | 1.8947(7) | | | | |
| $R_p$, $R_{wp}$, $R_e$, $\chi^2$ | 8.77 | 8.96 | 3.42 | 6.87 | | | | |
| Phase fraction | | 32% | | | | 68% | | |

Table 2 Structural parameter, fractional atomic coordinates and equivalent isotropic displacement parameter from the neutron refinements of the sample of KNN-50BNT at 300K. The isotropic displacement parameter $B$ is in Å$^2$

| x=0.50 at 300K | Tetragonal phase (P4bm) | | | | Cubic phase (Pm-3m) | | | |
|---|---|---|---|---|---|---|---|---|
| a, b, c (Å) | 5.5728(6) | 5.5728(6) | 3.9415(0) | | 3.9424(6) | 3.9424(6) | 3.9424(6) | |
| α, β, γ (°) | 90 | 90 | 90 | | 90 | 90 | 90 | |
| Na/K/Bi x, y, z B | 0 | ½ | 0.5564(1) | 3.9444(3) | 0 | 0 | 0 | 4.1251(6) |
| Nb/Ti x, y, z B | 0 | 0 | 0 | 0.6402(2) | ½ | ½ | ½ | 0.0676(7) |
| O1 x, y, z B | 0 | 0 | 0.5034(2) | 0.1981(5) | ½ | ½ | 0 | 2.9941(3) |
| O2 x, y, z B | 0.2327(3) | 0.2672(7) | 0.0057(1) | 0.4365(8) | | | | |
| $R_p$, $R_{wp}$, $R_e$, $\chi^2$ | 13.3 | 12.6 | 4.68 | 7.244 | | | | |
| Phase fraction | | 43% | | | | 57% | | |

Table 3 Fitting results according to the Vogel-Fulcher relation and the new glass model.

| Sample | | Glass model | | | New glass model | | | Ref |
|---|---|---|---|---|---|---|---|---|
| | | $\omega_{01}$ (Hz) | $E_{a1}$ (eV) | $T_f$ (K) | $\omega_{02}$ (Hz) | $E_{a2}$ (eV) | $p$ | |
| KNN-50BNT | $T'_m$-$\omega$ | 5.41×10$^{13}$ | 0.20 | 214.6 | | | | This work |
| | $T''_m$-$\omega$ | 8.92×10$^{11}$ | 0.18 | 117.4 | | | | |
| | $T'_m$-$\omega$ | | | | 5.40×10$^{12}$ | 0.069 | 3.26 | |
| | $T''_m$-$\omega$ | | | | 1.29×10$^9$ | 0.042 | 3.24 | |
| BaTi$_{0.7}$Sn$_{0.3}$O$_3$ | $T'_m$-$\omega$ | 2.0×10$^{11}$ | 0.042 | 110 | 8.0×10$^8$ | 0.017 | 7.03 | 27 |
| 0.9PMN-0.1PT | $T'_m$-$\omega$ | 4.5×10$^{15}$ | 0.15 | 205.1 | 1.2×10$^{13}$ | 0.037 | 8.81 | 25 |
| | $T''_m$-$\omega$ | 3.3×10$^{14}$ | 0.10 | 204.7 | | | | 28 |

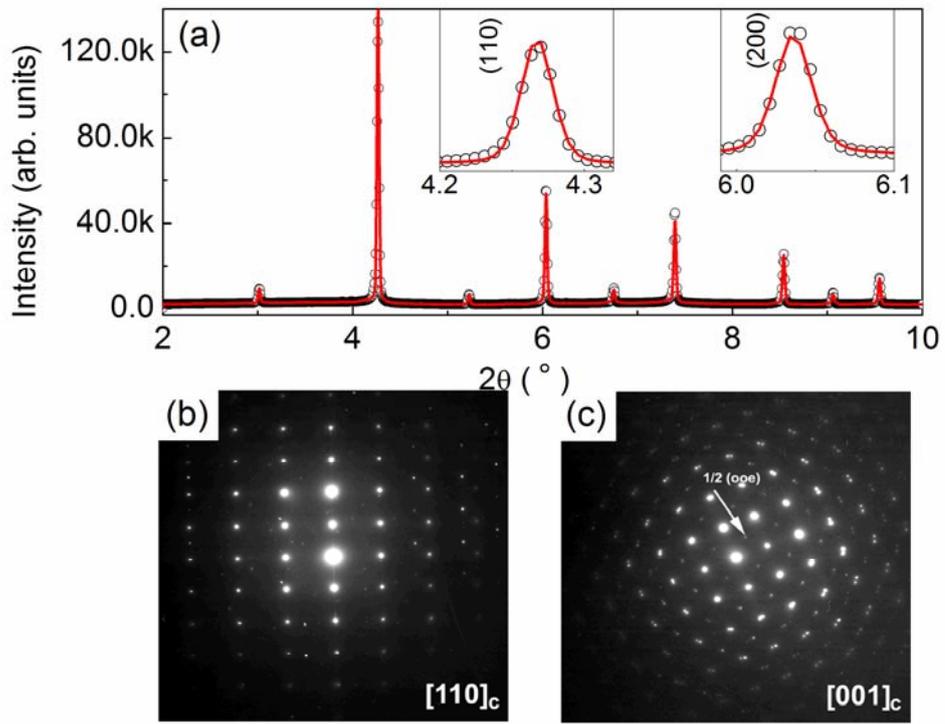

Fig. 1

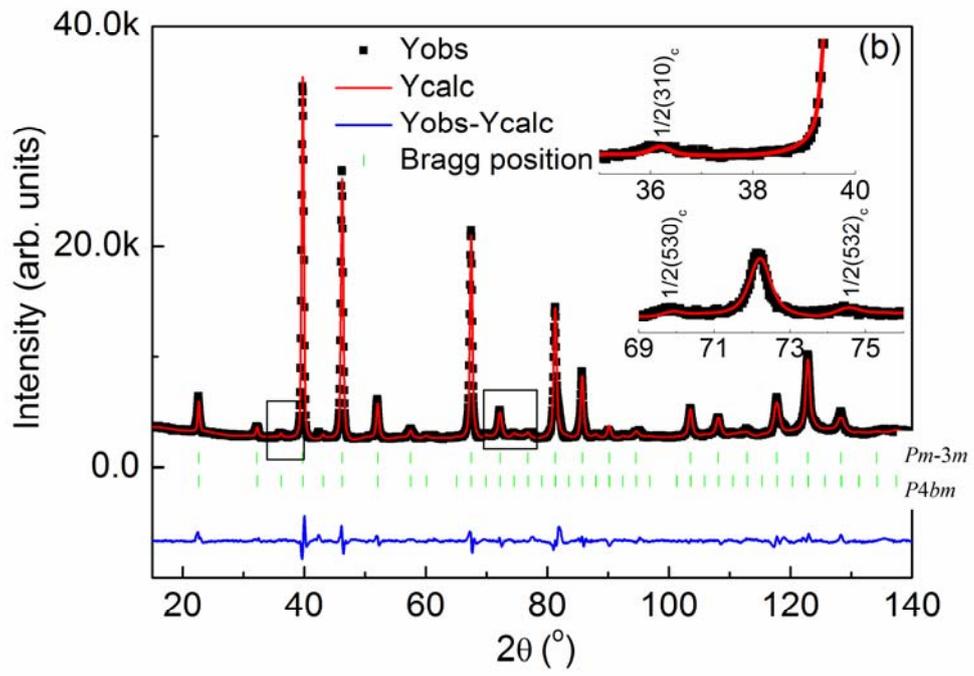

Fig. 2

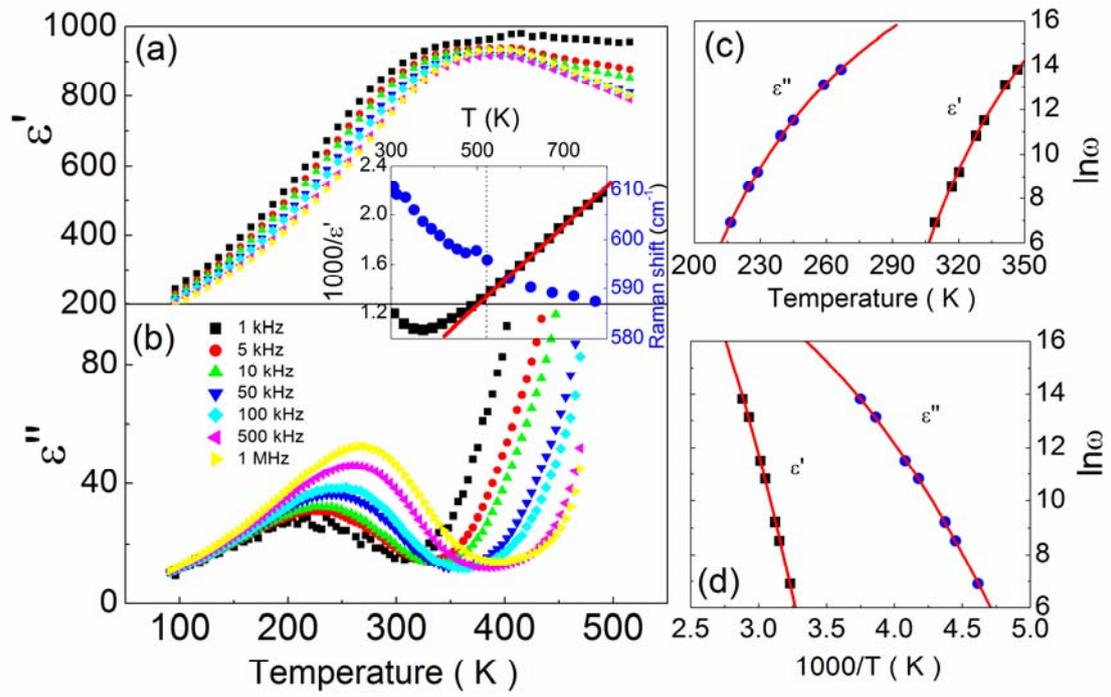

Fig. 3

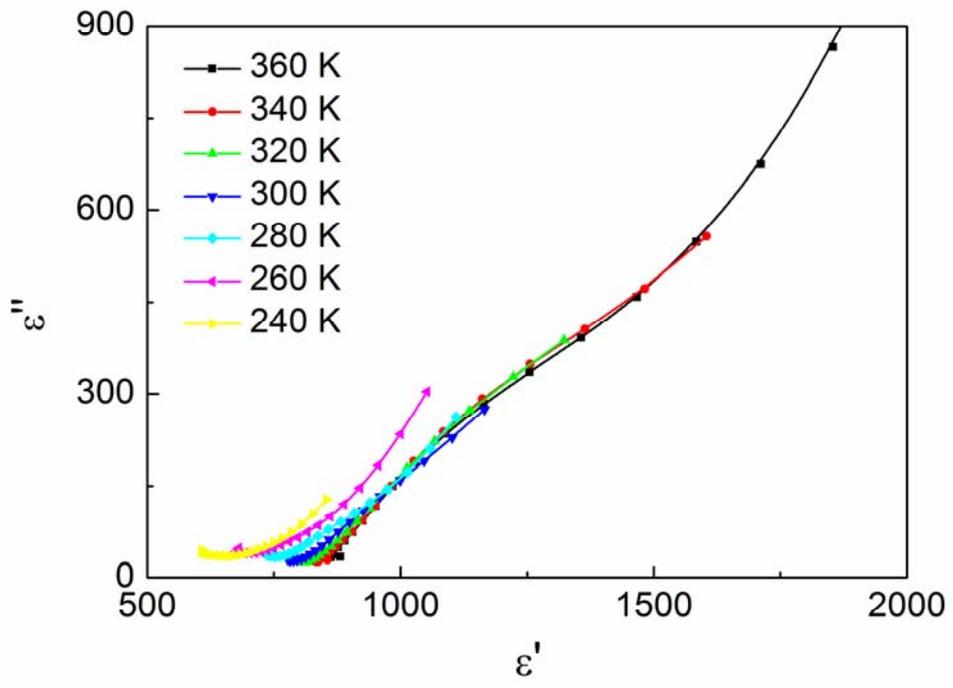

Fig. 4